# Ultrasound Characterizes Perfusion Trajectories in Gingival Wound Healing

Haowei Tai, PhD[1], Oscar Duran-Garnica, DDS, MS[2], Karen Villarreal, DDS, MSD[2], Hsun-Liang Chan, DDS, MS[2], Oliver D. Kripfgans, PhD[1, 3]

[1] Department of Radiology, University of Michigan School of Medicine, Ann Arbor, MI, USA

[2] Division of Periodontology, College of Dentistry, The Ohio State University, Columbus, OH, USA

[3] Department of Biomedical Engineering, College of Engineering, University of Michigan, Ann Arbor, MI, USA

**Corresponding author:** Oliver D. Kripfgans, University of Michigan School of Medicine, 1301 Catherine St., Medical Sciences Bldg 1, Room 6410, Ann Arbor, MI 48109-2026;

Email: greentom@umich.edu

**Abstract**

Monitoring wound healing after tooth extraction is important for identifying persistent inflammation and guiding clinical management. Current assessment relies primarily on visual examination, although wounds with similar appearances may have different underlying vascular remodeling. This study developed a quantitative ultrasound framework to measure longitudinal perfusion as a biomarker of tissue recovery and persistent inflammation. In this prospective observational study, 51 patients undergoing tooth extraction and socket preservation were imaged with color-flow ultrasound and Laser Speckle Contrast Analysis (LASCA) from approximately 2 weeks to 3 months after extraction. Acquired pre-scan-converted ultrasound data were processed for scan conversion, power estimation, and calculation of velocity-weighted color pixel density ($CPD_{vel}$). Gingival regions were automatically identified using BiSeNetV2 and evaluated using Dice similarity and Bland–Altman analysis. Perfusion was compared between healthy and inflamed visits during the early (<35 days) and recovery (≥35 days) periods and was also compared with LASCA measurements. Segmentation achieved a mean Dice score of 0.98 and a median Dice score of 0.99. Bland–Altman analysis showed minimal ROI area bias (−0.01 percentage points; 95% limits of agreement: −1.68 to 1.67). In healthy visits, median ultrasound $CPD_{vel}$ decreased by 81% from early to recovery ($p < 0.01$), while inflamed visits showed no significant change (26% reduction, $p = 0.16$). During recovery, perfusion remained higher in inflamed than healthy visits (0.12 vs 0.03 cm/s, $p = 0.04$). LASCA showed smaller temporal changes of 13% and 9%, respectively. Ultrasound perfusion correlated positively with LASCA in healthy visits ($\rho = 0.29$, $p = 0.03$), while inflamed visits showed a nonsignificant positive trend ($\rho = 0.28$, $p = 0.13$). Quantitative ultrasound enabled automated and reproducible gingival perfusion assessment with reduced sensitivity to scanner-dependent display settings. Distinct perfusion trajectories support its potential for longitudinal monitoring and detection of abnormal healing.

## Introduction

Inflammation is a fundamental biological response to tissue injury, infection, and surgical intervention that plays a central role in oral wound healing[1–3]. In the oral cavity, inflammatory changes commonly occur in periodontal tissues[4], peri-extraction wounds[5], and peri-implant sites[6], where they influence soft-tissue stability[7], wound repair[8], and subsequent bone healing[9]. Although a transient inflammatory response is essential for initiating tissue repair, excessive or persistent inflammation can delay wound closure, impair vascular maturation, disrupt collagen organization, and compromise regenerative outcomes[10]. Such prolonged inflammation has also been associated with delayed healing, increased risk of postoperative complications, and reduced treatment success[11]. Therefore, objective characterization of inflammation is important for determining whether oral tissues are recovering as expected after surgical injury.

A key physiological feature of inflammation and wound healing is alteration of the local microcirculation. Healthy oral soft tissues maintain tightly regulated vascular function that supports oxygen delivery, nutrient transport, immune surveillance, and tissue homeostasis[12]. Following tooth extraction, an immediate inflammatory response induces vasodilation and increased blood flow to deliver inflammatory cells and nutrients required for tissue repair. During subsequent healing, angiogenesis and vascular remodeling promote granulation tissue formation[13] and extracellular matrix deposition[14] before the vascular network gradually matures and returns toward homeostasis. In contrast, persistent inflammation prolongs vascular dilation, sustains inflammatory hyperemia, disrupts normal vascular remodeling, and delays tissue regeneration. Consequently, tissue perfusion changes dynamically throughout healing and closely reflects the physiological status of the tissue.

Characterizing longitudinal perfusion patterns holds promise for providing clinically meaningful information beyond visual tissue assessment, supporting timely clinical intervention and personalized management of wound healing[15]. During normal healing, perfusion is expected to decrease as inflammation resolves and vascular remodeling completes. Quantifying these changes may provide objective indicators of tissue recovery, support treatment planning, and enable earlier identification of impaired healing.

Current assessment of oral wound healing remains largely based on visual inspection[16], probing response[17], bleeding tendency[18], and clinician experience. Although practical and widely used, these methods primarily evaluate morphological changes and are inherently subjective. Their interpretation may vary with operator experience, local anatomy, lighting conditions, and patient-specific factors such as tissue pigmentation and baseline tissue characteristics. Moreover, visual examination provides mostly a qualitative assessment of tissue condition and does not directly reflect the underlying vascular activity driving healing. Tissues with similar clinical appearances may therefore have different underlying physiological states[19]. Objective quantification of tissue perfusion could improve diagnostic consistency, enable reproducible longitudinal monitoring across clinical visits[20], and provide measurable imaging biomarkers for evaluating healing progression and identifying patients who may benefit from additional clinical intervention.

Several optical imaging techniques have been investigated for assessing tissue perfusion, including capillaroscopy[21], intravital microscopy[22], laser Doppler flowmetry[23], and laser speckle contrast analysis (LASCA)[24]. These methods provide quantitative measures of superficial blood flow and capillary dynamics and have been used in dental and periodontal applications[25]. However, intraoral use is limited by shallow penetration, motion sensitivity, restricted posterior access, and probe-positioning challenges. Because they primarily assess superficial microvascular perfusion (<2 mm), they may not capture inflammatory and vascular changes occurring throughout the deeper gingival connective tissue.

Ultrasound imaging offers an attractive alternative for evaluating oral soft tissues because it is noninvasive, real-time, portable, and capable of cross-sectional imaging. The recent development of compact intraoral ultrasound probes has made ultrasound increasingly practical for imaging within the confined oral cavity[26–28]. Color-flow ultrasound measures blood flow from moving red blood cells and enables depth-resolved assessment of tissue vascularity and perfusion. Combined with anatomical imaging, it offers a promising approach for objective evaluation of oral inflammation and wound healing.

Despite these advantages, quantitative interpretation of color-flow ultrasound remains challenging because image appearance depends on scanner settings such as color gain and other system-dependent processing parameters. These settings are often adjusted during clinical examinations, which can confound longitudinal comparisons and differences between healthy and inflamed tissues. Clinically useful perfusion biomarkers therefore require quantitative analysis that minimizes dependence on operator-selected imaging settings and remains reproducible across examinations.

In this study, a signal-processing framework was developed to quantify local perfusion from complex-valued pre-scan-converted Total Recall Imaging (TRI) data while reducing sensitivity to system-dependent settings. Deep learning-based segmentation was incorporated to automatically identify the gingival region of interest. The framework was used to characterize longitudinal perfusion changes after tooth extraction and compare temporal patterns between healthy and inflamed tissues. It was hypothesized that perfusion would decrease during normal healing but remain elevated with persistent inflammation. These quantitative trajectories may provide objective markers for monitoring wound healing, treatment response, and abnormal healing.

## Materials and Methods

### Ultrasound Data Acquisition

This prospective longitudinal observational study included patients undergoing tooth extraction and socket preservation and was reported in accordance with the STROBE guidelines. The study was approved by the institutional review board, and all participants provided written informed consent in accordance with the Declaration of Helsinki. Patients undergoing tooth extraction and socket preservation were recruited at The Ohio State University between July 2024 and June 2026. Longitudinal pre-scan-converted complex-valued ultrasound data were collected from patients (n = 51) using a clinical ultrasound system (ZS3, Mindray, USA) equipped with a 128-element linear array transducer (L30-8, Mindray, USA). The transducer operated at a center frequency of 18 MHz in CSH24 mode with spatial compounding and harmonic imaging (12–24 MHz). During image acquisition, the clinician adjusted imaging parameters, including gain, velocity scale, imaging depth, and imaging plane, to optimize image quality[29,30]. Participants were scheduled for seven visits: V1, pre-procedure assessment; V2, tooth extraction and graft placement (day 0); and V3 to V7, postoperative follow-up (2 to 30 weeks). This study analyzed Visits V3–V6, which represent the longitudinal healing phase and typically span 2 weeks to 3 months after surgery. Inflammation status was determined by an experienced dentist (HC) based on tissue morphology (Appendix Methods). Visit 7 was excluded because it fell outside this analysis window and had limited imaging data.

### Ultrasound Data Preprocessing

Ultrasound TRI data containing B-mode and color flow information were processed offline.

Native polar-coordinate data were converted to a Cartesian grid through scan conversion[31] at a spatial resolution of 0.03*0.03 $mm^2$ per pixel to reconstruct the physical geometry of the imaged tissue. B-mode images were generated from the complex signal magnitude and log-compressed. Speckle was reduced using bilateral filtering[32], which smoothed local intensity variations while preserving tissue boundaries. Image intensities were clipped at the $1^{st}$ and $99^{th}$ percentiles, normalized, and adjusted using gamma and contrast correction to produce consistent grayscale images for visualization and tissue segmentation. Pixel-level velocity was calculated from the TRI-derived phase information using the velocity range reported by the ultrasound system. A fixed 17.5-dB power threshold, consistent with our related ultrasound analyses, was applied to exclude low-power pixels and unreliable velocity estimates. The remaining velocity data were then scan-converted according to the dimensions of the color flow box. Small, isolated regions and holes introduced during interpolation were removed using morphological processing. The retained velocity values were then mapped to the scanner color-flow scale and overlaid on the corresponding B-mode images.

**Gingival Tissue Identification**

B-mode frames were manually annotated by an experienced periodontist (OD) to delineate the gingival region of interest (ROI). Motion tracking based on corner detection and optical flow[33] was subsequently applied to propagate these annotations throughout each cine loop, substantially reducing manual labeling while generating image-segmentation pairs for every frame. The resulting dataset contained approximately 19,000 labeled ultrasound images. The image-label pairs were used to train a lightweight BiSeNetV2 real-time semantic segmentation network[34]. Segmentation performance was evaluated using the Dice similarity coefficient and Bland–Altman agreement between predicted and reference ROI areas. Training and test sets were separated at the patient level to ensure that images from the same patient were not included in both datasets (Appendix Methods). For each frame, velocity-weighted color pixel density ($CPD_{vel}$) was calculated as the sum of the absolute velocities of power-qualified pixels within the gingival ROI divided by the total number of pixels in the ROI.

**Optical Imaging**

Laser speckle contrast analysis (LASCA; PeriCam PSI, Perimed Inc., Sweden) was performed alongside ultrasound at each study visit as a reference measure of superficial tissue perfusion[35]. The system uses near-infrared laser illumination (785 nm) to generate two-dimensional surface perfusion maps by measuring temporal fluctuations in laser speckle patterns produced by moving blood cells. These fluctuations increase with blood flow, allowing LASCA to provide a noninvasive measure of superficial microvascular perfusion with an imaging depth of approximately 2 mm, a spatial resolution of ~10 μm, and frame rates of up to 113 frames/s. To account for limitations of LASCA, perfusion was normalized as the ratio of mean perfusion at the gingival extraction site to that of adjacent healthy tissue. Measurements were obtained at multiple postoperative time points from approximately 2 weeks to 3 months.

**Statistical Analysis**

Continuous variables were summarized according to their distributions. Visits were classified as healthy or inflamed and further grouped into early (<35 days) and recovery (≥35 days) periods. The 35-day cutoff was based on a prior case report that estimated perfusion recovery at 33.4 days after guided bone regeneration[36]. Bland–Altman analysis was used to evaluate agreement between predicted and reference ROI areas. One-sided Mann–Whitney U tests compared visit-level $\mathrm{CPD}_{vel}$ and normalized LASCA perfusion across healing periods and clinical-status groups, and one-sided Spearman rank correlations assessed their associations. Each visit was treated as an independent observation, and statistical significance was defined as $p<0.05$. Group comparisons and correlation analyses were not adjusted for potential confounders. Additional statistical details are provided in the Appendix.

## Results

Participant characteristics are summarized in Supplementary Table S1. Recruitment, follow-up completion, visit timing, and patient and visit counts for each analysis are provided in Supplementary Table S2. The exported TRI data were scan-converted to reconstruct the physical imaging geometry (**Figure 1**), which consisted of a 24° fan angle and a virtual apex located 30.1 mm above the top of the field of view, yielding maximum axial and lateral ranges of 15 and 18.78 mm, respectively. Images before and after scan conversion are shown for comparison. Bilateral filtering was applied to the reconstructed B-mode images to reduce speckle while preserving tissue boundaries. Finally, the color flow power map was used to exclude velocity pixels below a predefined threshold of 17.5 dB. The remaining pixels were then used to calculate $\mathrm{CPD}_{vel}$.

The first column of **Figure 2A** shows the predicted ROI boundaries overlaid on the corresponding B-mode images. The second column shows the pixel-wise probability maps, where red indicates a high probability of belonging to the ROI and blue indicates a lower probability, which was primarily observed near the tissue boundaries. The third column shows the uncertainty maps, in which red represents high uncertainty and blue represents high confidence. Additional segmentation results are provided in the Appendix Results. Segmentation performance was evaluated using 5700 gingival ROIs. The model achieved a mean Dice score of 0.98 and a median Dice score of 0.99 (IQR, 0.98–0.99). Most predictions exceeded the Dice = 0.95 reference line (98.8%). Greater variation was observed only for smaller ROIs. ROI size ranged from 0.19% to 16.01% of the image area, with a median size of 1.31% (**Figure 2B)**. The Bland–Altman plot in **Figure 2C** evaluated agreement between the predicted and annotated ROI areas after normalization to the total image area. The mean of the two normalized ROI areas is shown on the horizontal axis, while the predicted-minus- annotated difference is shown on the vertical axis. The mean bias was −0.01 percentage points, indicating negligible systematic underestimation by the model. The 95% limits of agreement ranged from −1.68 to 1.67 percentage points.

Representative longitudinal color-flow ultrasound images are shown in **Figure 3A**. Patients 48, 45, and 36 are presented from top to bottom, with Visits 3 to 6 arranged from left to right. Inflamed visits are outlined in red. Patient 48 was clinically healthy at all four visits. The ultrasound color-flow images showed a progressive decline in color-flow signals over time, with only minimal color-flow signals observed during the final two visits. Patient 45 exhibited inflammation only at Visit 4, while Visits 3, 5, and 6 were clinically healthy. Color-flow signal density increased at Visit 4 and decreased substantially during subsequent visits. Patient 36 exhibited inflammation at Visits 3 and 4 followed by healthy healing at Visits 5 and 6. High color flow signal density was observed during both inflamed visits and decreased during the subsequent healthy visits. Visual inspection of the color flow ultrasound images suggested a greater extent of color flow signals in inflamed tissue than in healthy tissue. **Figures 3B** and **3C** summarize the corresponding longitudinal $CPD_{vel}$ measurements and their distributions according to clinical status, with additional details provided in the appendix results. $CPD_{vel}$ was calculated for each frame within the segmented ROI and the median across all frames was used as the perfusion value for each visit.

A longitudinal comparison of clinical photographs, LASCA images, and ultrasound color-flow images for patient 37 across Visits 3 to 6 is presented in **Figure 4**. This patient was clinically inflamed at Visits 3 and 4 and classified as healthy at Visits 5 and 6. Similar temporal changes were observed in both LASCA and ultrasound color-flow images, with higher perfusion-related signals during inflamed visits and lower signals during healthy visits.

Temporal changes in ultrasound-derived $CPD_{vel}$ and LASCA perfusion were evaluated by comparing early (<35 days) and recovery (≥35 days) visits (**Figure 5**). Among healthy visits, median $CPD_{vel}$ decreased from 0.16 to 0.03 cm/s, corresponding to an 81% decrease and a recovery-to-early median ratio of 0.19 (one-sided Mann–Whitney U test, $p < 0.01$). Among inflamed visits, median $CPD_{vel}$ decreased from 0.16 to 0.12 cm/s, corresponding to a smaller decrease of 26% and a recovery-to-early median ratio of 0.74; this difference was not statistically significant ($p = 0.16$). LASCA perfusion showed a similar temporal pattern but with smaller changes than $CPD_{vel}$. Median LASCA perfusion decreased from 1.14 to 0.99 in healthy visits and from 1.52 to 1.38 in inflamed visits. These changes corresponded to decreases of 13% and 9%, with recovery-to-early median ratios of 0.87 and 0.91, respectively. Neither temporal difference reached statistical significance (healthy, $p = 0.13$; inflamed, $p = 0.48$). Healthy and inflamed visits were subsequently compared within each healing period. During the early period, median $CPD_{vel}$ was similar between healthy and inflamed visits (0.16 versus 0.16 cm/s, $p = 0.25$), as was median LASCA perfusion (1.14 versus 1.52, $p = 0.22$). During the recovery period, median $CPD_{vel}$ was higher in inflamed than in healthy visits (0.12 versus 0.03 cm/s, $p = 0.04$).

LASCA perfusion showed a corresponding difference, with a higher median in inflamed than in healthy visits (1.38 versus 0.99, $p = 0.02$). Exact-day-matched measurements showed weak positive associations between $CPD_{vel}$ and LASCA perfusion. The association met the prespecified significance threshold among healthy visits (Spearman's $\rho = 0.29$, one-sided $p = 0.03$) but not among inflamed visits ($\rho = 0.28$, one-sided $p = 0.13$).

**Discussion**

This study demonstrates the feasibility of using quantitative color flow ultrasound to objectively characterize longitudinal gingival tissue perfusion following tooth extraction and socket preservation. Healthy tissues exhibited progressive reduction in perfusion over time, consistent with vascular maturation and resolution of the inflammatory response, whereas inflamed tissues maintained relatively elevated perfusion throughout the healing period. These findings suggest that quantitative ultrasound can capture dynamic vascular changes associated with tissue healing and recovery. By providing objective measurements of longitudinal perfusion rather than qualitative assessment of vascularity, the proposed framework may improve evaluation of healing progression, facilitate assessment of treatment response, and provide quantitative biomarkers for identifying deviations from expected healing trajectories.

Unlike conventional color flow ultrasound, which is influenced by display settings such as color gain and threshold selection, the proposed framework extracts quantitative perfusion information before image display. As a result, the measurements are independent of these display settings except for the velocity scale, which can be read directly from the scanner and used for phase-to-velocity conversion. This eliminates dependence on other scanner-specific visualization parameters and improves comparability across longitudinal follow-up visits. Such reproducibility is essential if quantitative ultrasound is to become a practical tool for monitoring wound healing in routine dental practice.

Another important contribution of this study is the fully automated image analysis pipeline. Manual segmentation is labor intensive and introduces observer variability, particularly because the gingival tissue occupies only a small portion of the ultrasound image. This challenge becomes even greater for ultrasound cine loops, which often contain hundreds of frames, making manual segmentation impractical. To overcome this limitation, the BiSeNetV2 network was employed to automatically segment the gingival region before quantitative perfusion analysis. Bland–Altman analysis showed negligible average bias in ROI area relative to the total image area, suggesting that the model did not consistently overestimate or underestimate ROI size across the cohort. Therefore, segmentation errors were unlikely to introduce a consistent directional bias into the quantitative perfusion analysis.

Representative patient examples showed that changes in $CPD_{vel}$ paralleled changes in LASCA perfusion, with both measurements increasing during clinically inflamed visits and decreasing during clinically healthy visits. These findings are biologically consistent with the vascular response during wound healing. This pattern reflects the expected transition from inflammatory hyperemia to vascular maturation and tissue remodeling. The observed reduction in $CPD_{vel}$ during clinically healthy visits therefore reflects the expected physiological transition from active inflammation toward tissue recovery.

The longitudinal analysis further revealed distinct temporal characteristics between healthy and inflamed tissues. When visits were grouped by clinical status and healing period, $CPD_{vel}$ showed a marked decrease from the early to recovery period in clinically healthy tissue, whereas only a modest, nonsignificant decrease was observed in inflamed tissue. These observations suggest that successful wound healing is associated with progressive normalization of tissue vascularity, while persistent inflammation maintains increased blood flow over a longer period. Although additional studies with larger patient cohorts are required, these findings suggest that temporal perfusion patterns may help distinguish normal recovery from prolonged inflammation.

Comparison with LASCA provided additional validation of the proposed ultrasound measurements. Both modalities demonstrated higher perfusion in inflamed tissue than in healthy tissue during the recovery period, supporting the biological consistency of the measurements. However, ultrasound appeared more sensitive to longitudinal perfusion changes than LASCA. In healthy visits, ultrasound detected an 81% reduction in median perfusion between the early and recovery periods, whereas LASCA demonstrated only a 13% reduction. This difference might be partly due to methodological differences between the two imaging techniques. First, ultrasound quantified the absolute perfusion signal within the tissue, whereas LASCA reported the perfusion ratio between the extraction site and adjacent healthy tissue. Although normalization reduces intersubject variability, it may also reduce sensitivity if perfusion changes occur simultaneously in both the wound and the surrounding reference tissue. Second, ultrasound measures perfusion cross sectionally within deeper gingival tissue, whereas LASCA primarily measures superficial capillary perfusion near the epithelial surface because of its limited penetration depth (up to approximately 2 mm). Consequently, ultrasound may detect physiological changes occurring within deeper portions of the healing tissue that are less apparent using surface optical imaging alone[37,38]. After stratification by inflammation status, a significant positive correlation between ultrasound-derived $CPD_{vel}$ and LASCA was observed in the healthy group, whereas the inflamed group demonstrated a similar positive trend that did not reach statistical significance. It is also important to recognize that ultrasound and LASCA assess tissue perfusion using fundamentally different physical principles. LASCA estimates superficial tissue perfusion by analyzing laser speckle fluctuations generated by moving red blood cells, whereas ultrasound estimates perfusion from moving blood scatterers within the imaging volume[39]. LASCA is primarily sensitive to the superficial epithelium, where vascular density is relatively low[40], whereas ultrasound penetrates deeper to interrogate larger feeding vessels and vascular networks within the deep gingival tissues. Consequently, perfect agreement between the two techniques is not expected. Nevertheless, the positive associations observed in both groups suggest that the proposed ultrasound framework captures physiologically meaningful changes in tissue perfusion while providing complementary information beyond that available from surface optical imaging.

Several limitations should be considered when interpreting these findings. First, although the proposed segmentation model achieved excellent performance, comparisons with alternative deep learning architectures were not performed. Such comparisons could help determine whether the selected network provides the best balance of accuracy, computational efficiency, and robustness. Segmentation agreement was also evaluated at the frame level, with multiple frames obtained from the same cine loop. The Bland–Altman analysis should therefore be interpreted as a descriptive assessment within the test dataset rather than a patient-level estimate. Second, this study included a relatively small number of patients from a single clinical center. Larger prospective multicenter studies are needed to confirm these findings and assess generalizability across patient populations, imaging conditions, and operators. One-sided tests were used for prespecified directional hypotheses. However, the findings should be considered exploratory because the analyses did not account for within-patient correlation or potential confounding. Selection based on follow-up data availability may also limit generalizability. Future studies should use mixed-effects models and repeated-measures agreement analyses to account for correlated observations. Third, the current analysis focused on $CPD_{vel}$ as the primary perfusion biomarker. Additional ultrasound-derived parameters, including power-based measurements and multiparametric quantitative ultrasound features, may provide complementary information about vascular physiology and tissue condition. Finally, although LASCA served as an important reference modality, additional physiological measurements, histological validation, and clinical outcomes are needed to strengthen biological interpretation and clinical utility. Inflammation status was also determined by routine visual clinical assessment. Future studies incorporating more objective measures of inflammation could improve classification of healing status and provide a stronger reference for interpreting ultrasound perfusion changes.

In conclusion, quantitative ultrasound provided an automated method for assessing gingival perfusion during post-extraction healing. Deep learning-based segmentation and $CPD_{vel}$ derived from color-flow TRI data enabled objective longitudinal assessment and revealed distinct perfusion trajectories in healthy and inflamed tissues. The greater dynamic range than LASCA suggests that subsurface ultrasound perfusion imaging may provide complementary information for monitoring tissue recovery, treatment response, and abnormal healing.

## Author Contributions

H. Tai contributed to conception and design, data analysis and interpretation, drafted and critically revised the manuscript; O. Duran-Garnica and K. Villarreal contributed to data acquisition; H.-L. Chan and O.D. Kripfgans contributed to conception and design, data acquisition, analysis, and interpretation, and critically revised the manuscript. All authors gave their final approval and agreed to be accountable for all aspects of the work, ensuring its integrity and accuracy. The funder had no role in study design, data collection or analysis, manuscript preparation, or the decision to submit the manuscript for publication.

**Data availability statement**

All data supporting the findings of this study are included in the published article. The raw data underlying the figures, tables, and quantitative analyses, together with the associated spreadsheets, source files, and analysis code, have been deposited in Figshare and are publicly available at: https://doi.org/10.6084/m9.figshare.34004112.

**Acknowledgement**

This research was supported by the National Institutes of Health (NIH) R01DE030872.

**Declaration of Conflicting Interests**

The authors declared no potential conflicts of interest with respect to the research, authorship, and/or publication of this article.

**References**

1. Baseri M, Radmand F, Hamedi R, et al. Immunological Aspects of Dental Implant Rejection. *BioMed Research International*. 2020;2020(1):7279509.

2. Boynueğri D, Nemli SK, Kasko YA. Significance of keratinized mucosa around dental implants: a prospective comparative study. *Clinical Oral Implants Research*. 2013;24(8):928–933.

3. Noronha Oliveira M, Schunemann WVH, Mathew MT, et al. Can degradation products released from dental implants affect peri-implant tissues? *Journal of Periodontal Research*. 2018;53(1):1–11.

4. Loos BG, Van Dyke TE. The role of inflammation and genetics in periodontal disease. *Periodontology 2000*. 2020;83(1):26–39.

5. Seigneur M, Cloitre A, Malard O, et al. Teeth roots displacement in the maxillary sinus: characteristics and management. *J Oral Med Oral Surg*. 2020;26(3):34.

6. Salvi GE, Stähli A, Imber J-C, et al. Physiopathology of peri-implant diseases. *Clinical Implant Dentistry and Related Research*. 2023;25(4):629–639.

7. Kim JC, Lee M, Yeo I-SL. Three interfaces of the dental implant system and their clinical effects on hard and soft tissues. *Mater. Horiz.* 2022;9(5):1387–1411.

8. Chen Y, Gao B, Cai W, et al. Oral mucosa: anti-inflammatory function, mechanisms, and applications. *J. Mater. Chem. B*. 2025;13(13):4059–4072.

9. Thomas MV, Puleo DA. Infection, Inflammation, and Bone Regeneration: a Paradoxical Relationship. *J Dent Res*. 2011;90(9):1052–1061.

10. Vaz MM, Lopes LG, Cardoso PC, et al. Inflammatory response of human dental pulp to at-home and in-office tooth bleaching. *J. Appl. Oral Sci.* 2016;24:509–517.

11. D'Aiuto F, Parkar M, Andreou G, et al. Periodontitis and Systemic Inflammation: Control of the Local Infection is Associated with a Reduction in Serum Inflammatory Markers. *J Dent Res*. 2004;83(2):156–160.

12. Wahyudi D, Kusumasari C. Oxygen as obturation biomaterial in endodontic treatment: development of novel membranous dental restoration system. *F1000Res*. 2024;12:380.

13. Sculean A, Gruber R, Bosshardt DD. Soft tissue wound healing around teeth and dental implants. *Journal of Clinical Periodontology*. 2014;41(s15):S6–S22.

14. Goldberg M, Smith AJ. Cells and Extracellular Matrices of Dentin and Pulp: A Biological Basis for Repair and Tissue Engineering. *Critical Reviews in Oral Biology & Medicine*. 2004;15(1):13–27.

15. Tavelli L, Nguyen T, Rodriguez MV, et al. Tissue Perfusion and Biomarkers Assessment Following Root Coverage Procedures. *Journal of Periodontal Research*. 2025;60(11):1117–1131.

16. Wenzel A, Larsen MJ, Feierskov O. Detection of Occlusal Caries without Cavitation by Visual Inspection, Film Radiographs, Xeroradiographs, and Digitized Radiographs. *Caries Res*. 2009;25(5):365–371.

17. Fernandes GVO, Muller R. Back to the Basics: Mastering Probing Techniques for Comprehensive Periodontal Assessment. *The Journal of Basic and Clinical Dentistry*. 2024;1(1):1–8.

18. Mohamed Rohani M, Fisal AA. The Dental Management of Patients With Common Hematological Disorders and Bleeding Tendency. *Special Care in Dentistry*. 2025;45(4):e70090.

19. Alves CH, Russi KL, Rocha NC, et al. Host-microbiome interactions regarding peri-implantitis and dental implant loss. *J Transl Med*. 2022;20(1):425.

20. Katz MS, Ooms M, Winnand P, et al. Evaluation of perfusion parameters of gingival inflammation using laser Doppler flowmetry and tissue spectrophotometry– a prospective comparative clinical study. *BMC Oral Health*. 2023;23(1):761.

21. Scardina GA, Cacioppo A, Messina P. Periodontal microcirculation in diabetics: An in vivo non-invasive analysis by means of videocapillaroscopy. *Med Sci Monit*. 2012;18(2):CR58–CR64.

22. Yang D, Ortinau L, Jeong Y, et al. Advances and challenges in intravital imaging of craniofacial and dental progenitor cells. *genesis*. 2022;60(8–9):e23498.

23. Orekhova LY, Barmasheva AA. Doppler flowmetry as a tool of predictive, preventive and personalised dentistry. *EPMA Journal*. 2013;4(1):21.

24. Rodriguez A, Kripfgans O, Aellos F, et al. Non-invasive and quantitative methods for assessment of blood flow in periodontal and oral soft tissues: a systematic review. *Front. Dent. Med.* 2025;6.

25. Ghouth N, Duggal MS, BaniHani A, et al. The diagnostic accuracy of laser Doppler flowmetry in assessing pulp blood flow in permanent teeth: A systematic review. *Dental Traumatology*. 2018;34(5):311–319.

26. Rodriguez Betancourt A, Samal A, Chan H-L, et al. Overview of Ultrasound in Dentistry for Advancing Research Methodology and Patient Care Quality with Emphasis on Periodontal/Peri-implant Applications. *Zeitschrift für Medizinische Physik*. 2023;33(3):336–386.

27. Marotti J, Heger S, Tinschert J, et al. Recent advances of ultrasound imaging in dentistry – a review of the literature. *Oral Surgery, Oral Medicine, Oral Pathology and Oral Radiology*. 2013;115(6):819–832.

28. Elbarbary M, Sgro A, Khazaei S, et al. The applications of ultrasound, and ultrasonography in dentistry: a scoping review of the literature. *Clin Oral Invest*. 2022;26(3):2299–2316.

29. Tattan M, Sinjab K, Lee E, et al. Ultrasonography for chairside evaluation of periodontal structures: A pilot study. *Journal of Periodontology*. 2020;91(7):890–899.

30. Kripfgans OD, Chan H-L. Preliminary Experience in Transducer Preparation for Intraoral Imaging. *Journal of Ultrasound in Medicine*. 2023;42(9):2155–2166.

31. Tai H, Basavarajappa L, Hoyt K. 3-D H-scan ultrasound imaging of relative scatterer size using a matrix array transducer and sparse random aperture compounding. *Computers in biology and medicine*. 2022;151:106316.

32. Singh K, Sharma B, Singh J, et al. Local Statistics-based Speckle Reducing Bilateral Filter for Medical Ultrasound Images. *Mobile Netw Appl*. 2020;25(6):2367–2389.

33. Ayvali E, Desai JP. Optical Flow-Based Tracking of Needles and Needle-Tip Localization Using Circular Hough Transform in Ultrasound Images. *Ann Biomed Eng*. 2015;43(8):1828–1840.

34. Yu C, Gao C, Wang J, et al. BiSeNet V2: Bilateral Network with Guided Aggregation for Real-Time Semantic Segmentation. *Int J Comput Vis*. 2021;129(11):3051–3068.

35. Gnyawali SC, Blum K, Pal D, et al. Retooling Laser Speckle Contrast Analysis Algorithm to Enhance Non-Invasive High Resolution Laser Speckle Functional Imaging of Cutaneous Microcirculation. *Sci Rep*. 2017;7(1):41048.

36. Durán-Garnica O, Villarreal-Arizpe K, Rodríguez A, et al. Flap perfusion of superficial split flap design for guided bone regeneration: A case report with ultrasonography. *Clinical Advances in Periodontics*. 2026;16(3):665–675.

37. Tai H, Dolui S, Khairalseed M, et al. Three-dimensional voxel-level classification of ultrasound scattering. In: *2020 IEEE 17th International Symposium on Biomedical Imaging (ISBI)*. IEEE; 2020:1838–1841.

38. Tai H, Worlikar T, Xu Z, et al. Quantification of Histotripsy Dosage Using Machine Learning-Enhanced Ultrasound Imaging Analysis: Correlation with Histological Outcomes. *Academic Radiology*. 2026.

39. Yiu BYS, Yu ACH. High-Frame-Rate Ultrasound Color-Encoded Speckle Imaging of Complex Flow Dynamics. *Ultrasound in Medicine & Biology*. 2013;39(6):1015–1025.

40. Aharinejad S, Lametschwandtner A, Franz P, et al. The Vascularization of the Digestive Tract Studied by Scanning Electron Microscopy with Special Emphasis on the Teeth, Esophagus, Stomach, Small and Large Intestine, Pancreas, and Liver. *Scanning Microscopy*. 1991;5(3).

## Figure Legends

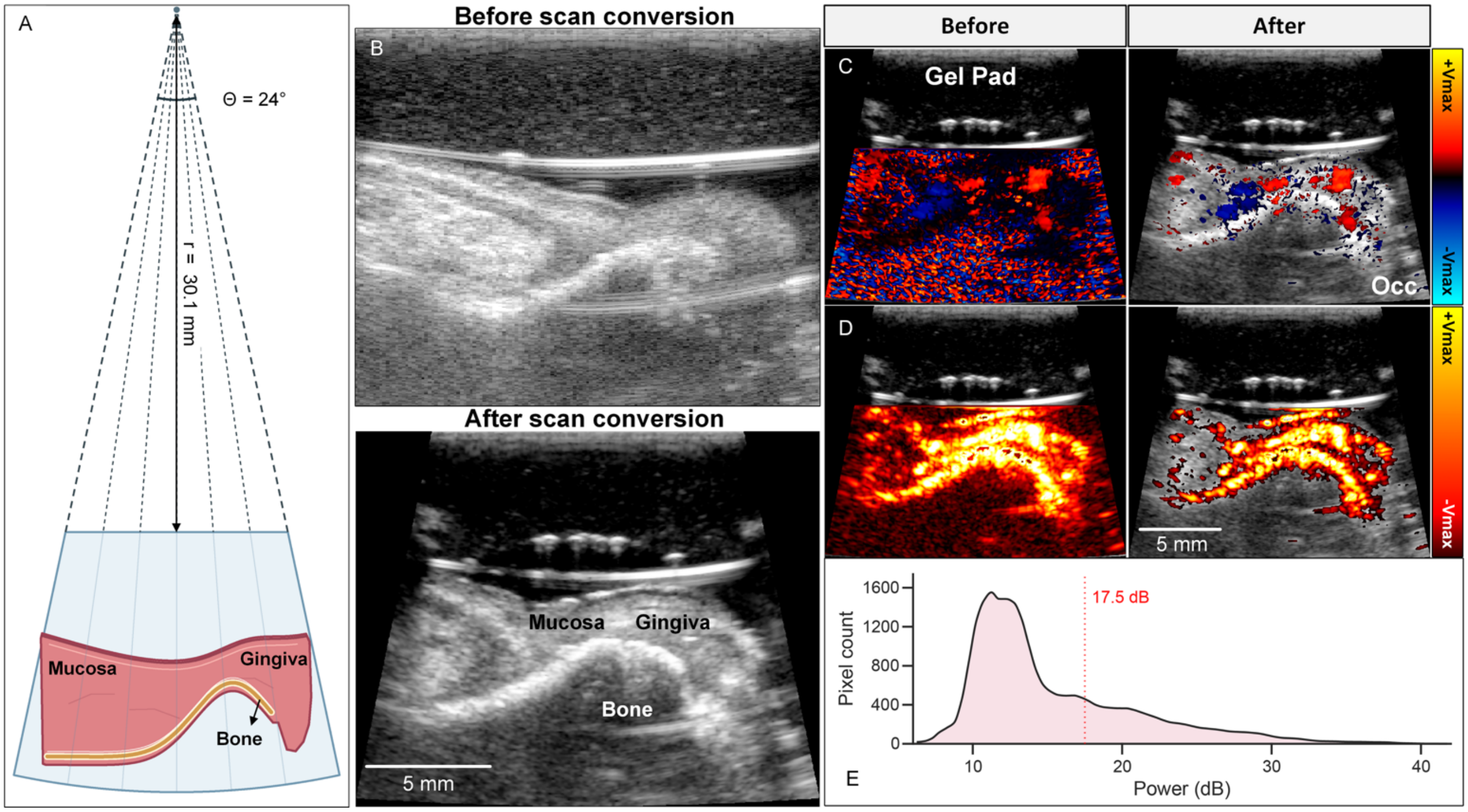

**Figure 1.** Workflow of the proposed quantitative ultrasound perfusion analysis. (A) Pre-scan-converted ultrasound data were scan converted according to the scan-line angles reported by the ultrasound system to reconstruct the correct image geometry and produce the corresponding B-mode image. (B) A bilateral filter was applied to reduce speckle noise while preserving tissue boundaries. (C) The phase information embedded in the Total Recall Imaging (TRI) data was used to estimate tissue velocity. (D) Tissue power was calculated. (E) A 17.5 dB power threshold was applied to suppress low-power noise before producing the final $CPD_{vel}$ color-flow image.

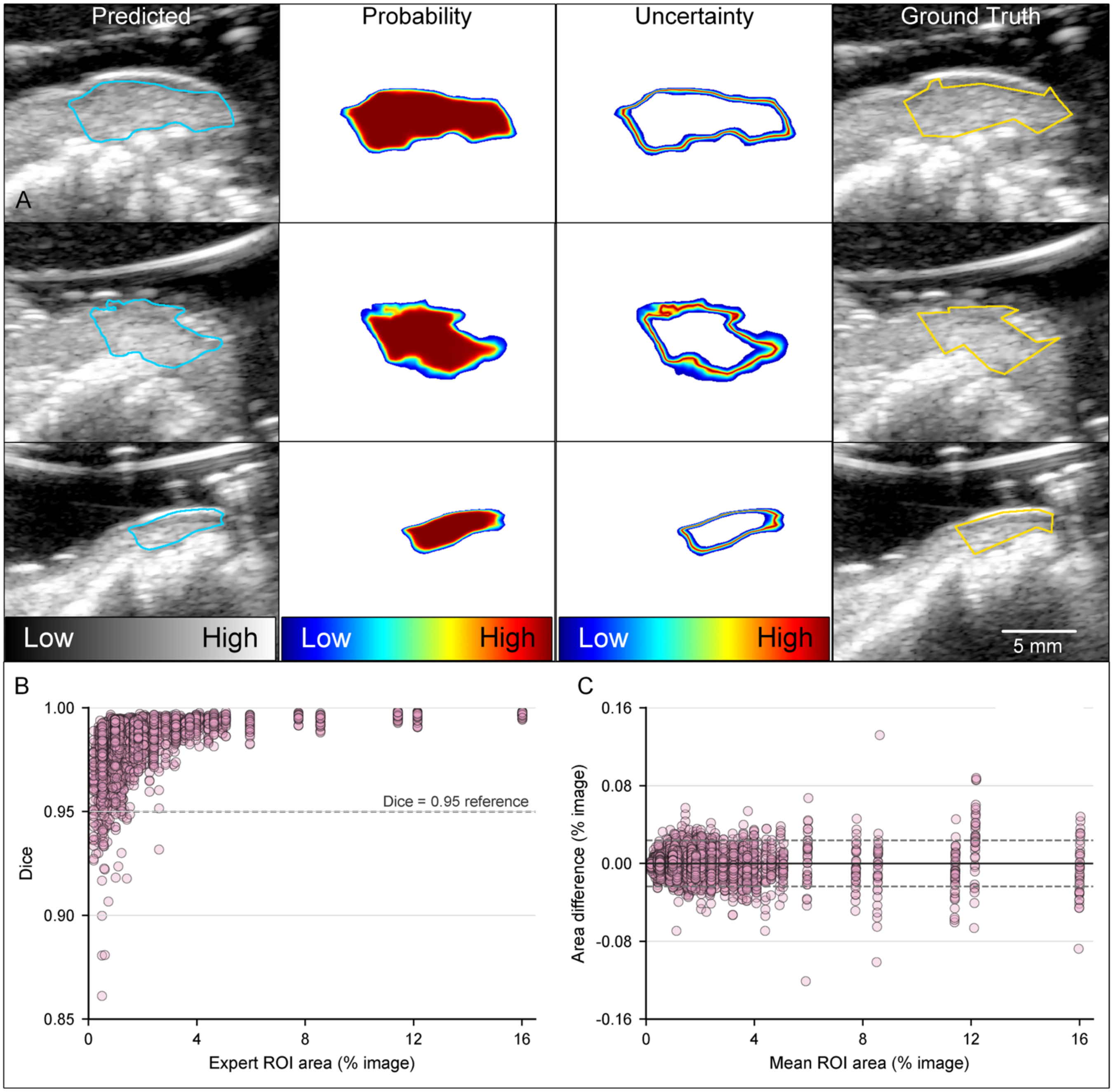

**Figure 2.** Evaluation of the automated gingival segmentation model. (A) Representative segmentation results showing the predicted gingival region of interest (ROI), the corresponding pixel-wise probability map, and the uncertainty map generated by the deep learning model. The probability and uncertainty maps provide a visual interpretation of the model predictions and confidence. (B) Dice similarity coefficient of each segmented ROI plotted as a function of ROI

size relative to the entire image. (C) Bland–Altman analysis of agreement between ROI areas predicted by the segmentation model and those obtained from expert manual annotations. The horizontal axis represents the mean predicted and manual ROI area as a percentage of the total image area, calculated as $100 * (A_{pred} + A_{manual})/2N$. The vertical axis represents the predicted-minus-manual ROI area difference relative to the total image area, calculated as $100 * (A_{pred} - A_{manual})/N$. Positive values indicate model overestimation and negative values indicate underestimation. The solid line represents the mean bias, and the dashed lines represent the 95% limits of agreement.

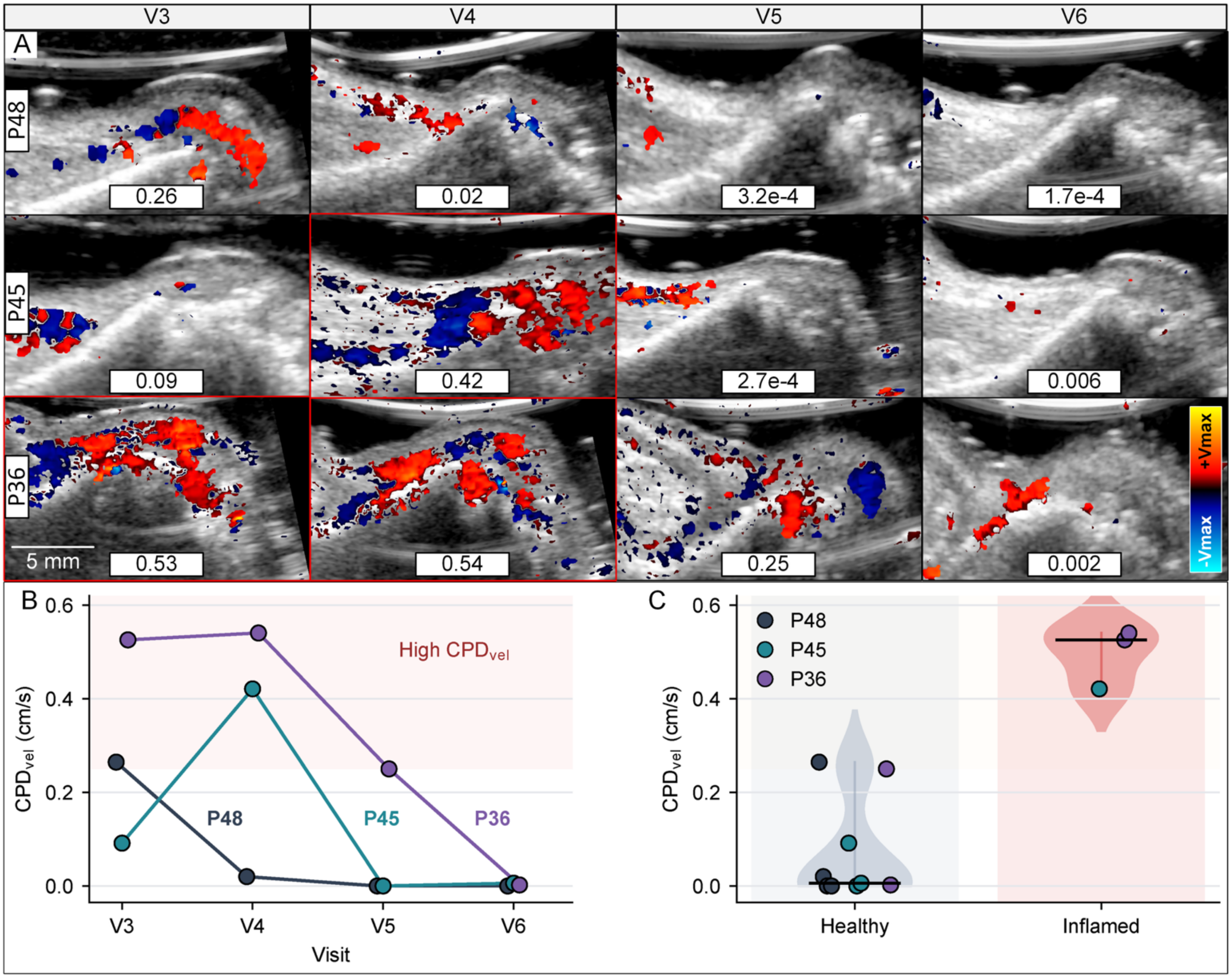

**Figure 3.** Representative longitudinal color-flow ultrasound images and quantitative perfusion measurements. (A) Representative color-flow ultrasound images from three patients acquired during longitudinal follow-up. Patients 48, 45, and 36 are shown from top to bottom, and Visits 3–6 are shown from left to right. Red borders indicate visits clinically classified as inflamed. Patient 48 remained clinically healthy throughout follow-up and showed progressively reduced color-flow signals over time. Patient 45 exhibited inflammation only at Visit 4, with increased color-flow signals during the inflamed visit that decreased during subsequent healthy visits. Patient 36 was classified as inflamed at Visits 3 and 4 and healthy at Visits 5 and 6, with prominent color-flow signals during the inflamed visits and reduced signals during healing. (B) Longitudinal $CPD_{vel}$ measurements for each patient. For each visit, $CPD_{vel}$ was calculated for each frame and summarized across frames using the median and interquartile range. Values are reported in cm/s without logarithmic transformation. (C) Distribution of $CPD_{vel}$ values by clinical status for the representative patients shown in panel (A).

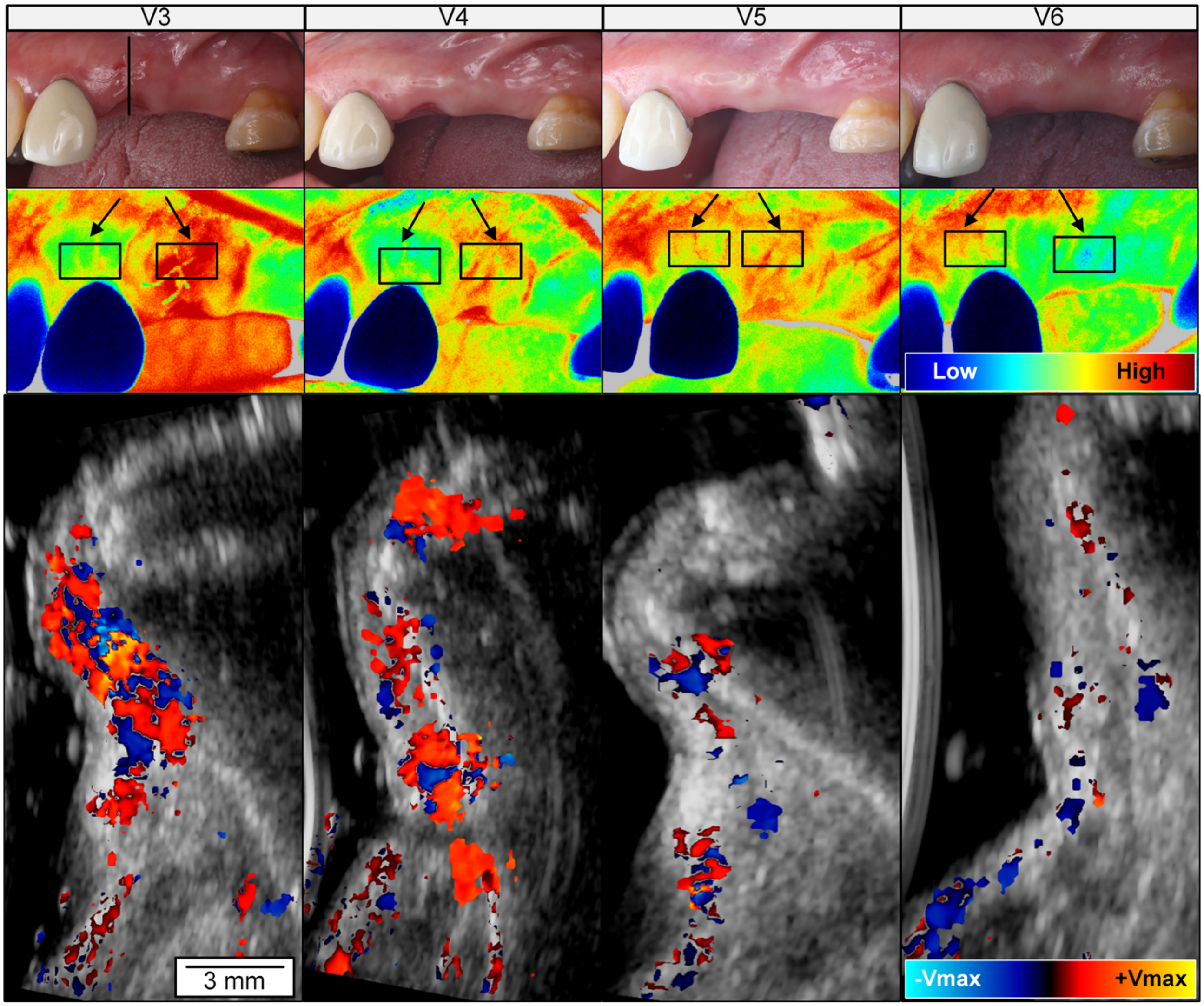

**Figure 4.** Comparison of clinical photographs, LASCA images, and ultrasound color flow images during longitudinal healing in patient 37. Clinical photographs (top row), corresponding LASCA images (middle row), and ultrasound color flow images (bottom row) are shown for Visits 3 to 6 from left to right. Visits 3 and 4 were clinically classified as inflamed, whereas Visits 5 and 6 were classified as healthy. The black line in the clinical photographs indicates the estimated ultrasound imaging region over the gingival soft tissue. The corresponding LASCA and ultrasound images demonstrate higher perfusion-related signals during the inflamed visits and reduced signals during the later healthy visits.

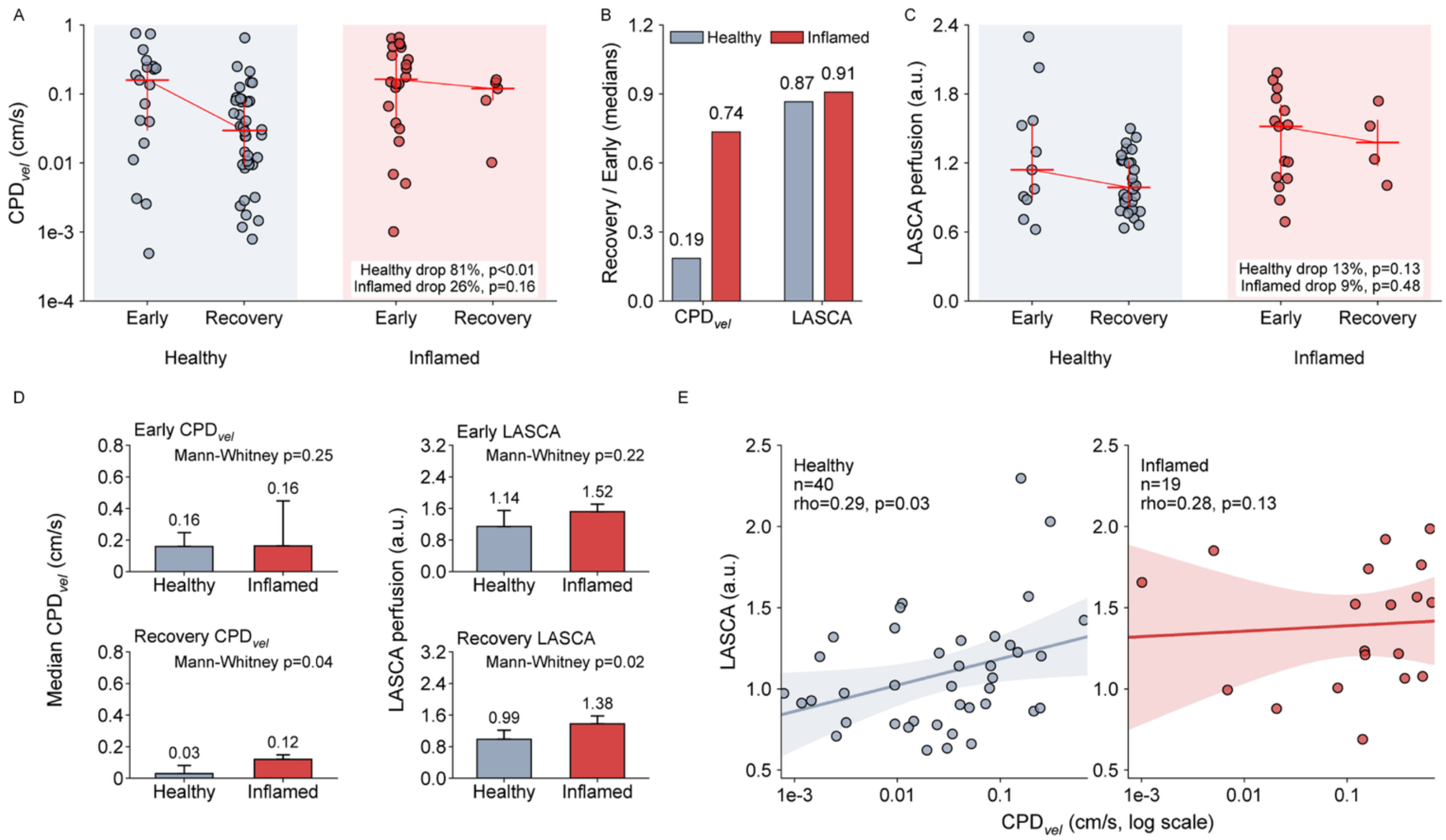

**Figure 5.** Temporal comparison of ultrasound-derived $CPD_{vel}$ and normalized LASCA perfusion during early (<35 days) and recovery (≥35 days) visits. (A) Distribution of $CPD_{vel}$ for healthy and inflamed visits. Each point represents one visit-level value defined as the median across frames within the corresponding cine loop. Values are shown in cm/s on a base-10 logarithmic y-axis. Horizontal lines and whiskers indicate the median and interquartile range (IQR). (B) Recovery-to-early median ratios for $CPD_{vel}$ and LASCA. Values below 1 indicate lower perfusion during recovery. (C) Distribution of normalized LASCA perfusion during the early and recovery periods, showing a similar pattern to $CPD_{vel}$ but smaller temporal changes in both healthy and inflamed visits. (D) $CPD_{vel}$ and LASCA measurements for healthy and inflamed visits within each period. Bars represent the median, and upper whiskers extend to the third quartile (Q3). (E) Correlation between exact-day-matched $CPD_{vel}$ and normalized LASCA measurements for healthy and inflamed visits. Spearman correlations were calculated using untransformed values. $CPD_{vel}$ is shown in cm/s on a base-10 logarithmic x-axis. For visualization, ordinary least-squares fits were calculated using log10-transformed $CPD_{vel}$, with shaded 95% confidence intervals for the fitted mean. Panels A and C used one-sided Mann–Whitney U tests for the prespecified hypothesis of higher perfusion during the early period. Panel D used one-sided Mann–Whitney U tests for higher perfusion in inflamed visits. Panel E used one-sided Spearman tests for a positive association. $CPD_{vel}$ values of 0 were excluded before analysis and plotting.